
\input phyzzx

\tolerance=10000
\hfuzz=5pt

\let\refmark=\NPrefmark  
\def\define#1#2\par{\def#1{\Ref#1{#2}\edef#1
{\noexpand\refmark{#1}}}}
\def\con#1#2\noc{\let\?=\Ref\let\<=\refmark\let\Ref=\REFS
         \let\refmark=\undefined#1\let\Ref=\REFSCON#2
         \let\Ref=\?\let\refmark=\<\refsend}

\define\RSSA
J. Schwarz and A. Sen, ``Duality Symmetric Actions,''
preprint NSF-ITP-93-46 (hep-th/9304154).

\define\RSSB
J. Schwarz and A. Sen,  ``Duality Symmetries of 4D Heterotic
Strings,'' preprint NSF-ITP-93-64 (hep-th/9305185).

\define\RKUGO
T. Kugo and B. Zwiebach, Prog. Theor. Phys. {\bf 87} (1992) 801.

\define\RTOWN
J. Azcarraga, J. Gauntlett, J. Izquierdo and
P. Townsend, Phys. Rev. Lett. {\bf 63} (1989) 2443.

\define\RNEPO
R. Nepomechie, Phys. Rev. {\bf D31} (1984) 1921;
C. Teitelboim, Phys. Lett. {\bf B167} (1986) 69.

\define\RRECENT
R. Kallosh and T. Ortin, preprint SU-ITP-93-3 (hep-th/9302109);
P. Bin\'etruy, preprint NSF-ITP-93-60;
M. Duff and R. Khuri, preprint
CTP/TAMU-17/93 (hep-th/9305142).

\define\RFSSS
A. Font, L. Ib\'a\~nez, D. Lust and F. Quevedo,
Phys. Lett. {\bf B249} (1990)
35;
S.J. Rey, Phys. Rev. {\bf D43} (1991) 526;
A. Shapere, S. Trivedi and F. Wilczek, Mod. Phys. Lett. {\bf A6}
(1991) 2677.

\define\RCECO
S. Cecotti, S. Ferrara and L. Girardello,
Nucl. Phys. {\bf B308} (1988)
436;
M. Duff, Nucl. Phys. {\bf B335} (1990) 610;
J. Molera and B. Ovrut, Phys. Rev. {\bf D40} (1989) 1146;
T. Kugo and B. Zwiebach, Prog. Theor. Phys. {\bf 87} (1992) 801.

\define\RNARAIN
K. Narain, Phys. Lett. {\bf B169} (1986) 41;
K. Narain, H. Sarmadi and E. Witten, Nucl. Phys. {\bf B279} (1987) 369.

\define\ROLD
E. Cremmer, J. Scherk and S. Ferrara,
Phys. Lett. {\bf B74} (1978) 61;
M. Gaillard and B. Zumino, Nucl. Phys. {\bf B193} (1981) 221;
M. De Roo, Nucl. Phys. {\bf B255} (1985) 515.

\define\RJACKIW
R. Floreanini and R. Jackiw, Phys. Rev. Lett. {\bf 59} (1987) 1873.

\define\RSLTZ
A. Sen, preprint TIFR-TH-93-03 (hep-th/9302038),
to appear in Int. J. Mod. Phys. A.

\define\ROLIVE
C. Montonen and D. Olive, Phys. Lett. {\bf B72} (1977) 117;
P. Goddard, J. Nuyts and D. Olive, Nucl. Phys. {\bf B125} (1977) 1;
H. Osborn, Phys. Lett. {\bf B83} (1979) 321.

\define\RDUALSUGRA
A. Chamseddine, Phys. Rev. {\bf D24} (1981) 3065;
S. Gates and H. Nishino, Phys. Lett. {\bf B173} (1986) 52;
A. Salam and E. Sezgin, Physica Scripta {\bf 32} (1985) 283.

\define\RHENN
M. Henneaux and C. Teitelboim, p. 79 in Proc. {\it Quantum Mechanics of
Fundamental Systems 2} (Santiago 1987); Phys. Lett. {\bf B206} (1988) 650.

\define\RHARLIU
J. Harvey and J. Liu, Phys. Lett. {\bf B268} (1991) 40;
R. Kallosh, A.
Linde, T. Ortin, A. Peet and A. Van Proeyen, Phys. Rev. {\bf D46} (1992)
5278; T. Ortin, preprint SU-ITP-92-24 (hep-th/9208078).

\define\RSTROM
A. Strominger, Nucl. Phys. {\bf B343} (1990) 167; C. Callan, J. Harvey and
A. Strominger, Nucl. Phys. {\bf B359} (1991) 611;
{\bf B367} (1991) 60;
preprint EFI-91-66 (hep-th/9112030).

\define\RDUFF
M. Duff, Class. Quantum Grav. {\bf 5} (1988) 189;
M. Duff and J. Lu, Nucl. Phys. {\bf B354} (1991) 129, 141;
{\bf B357} (1991) 534;
Phys. Rev. Lett. {\bf 66} (1991) 1402;
Class. Quantum Grav. {\bf 9} (1991) 1;
M. Duff, R. Khuri and J. Lu, Nucl. Phys. {\bf B377} (1992) 281;
J. Dixon, M. Duff and J. Plefka, Phys. Rev. Lett. {\bf 69} (1992) 3009.

\define\RMAHSCH
J. Maharana and J. Schwarz, Nucl. Phys. {\bf B390} (1993) 3.

\define\RSCHWARZ
J. Schwarz, preprint CALT-68-1815 (hep-th/9209125).

\define\RDUALITY
A. Sen, preprint TIFR-TH-92-41 (hep-th/9207053)
(to appear in Nucl.
Phys. B).

\define\RDIRAC
P. Dirac, Proc. R. Soc. {\bf A133} (1931) 60; J. Schwinger, Phys. Rev.
{\bf 144} (1966) 1087; {\bf 173} (1968) 1536; D. Zwanziger, Phys. Rev. {\bf
176} (1968) 1480, 1489;
E. Witten, Phys. Lett. {\bf 86B} (1979) 283.

\define\RDYONO
A. Sen, Phys. Lett. {\bf B303} (1993) 22;
A. Sen, preprint NSF-ITP-93-29 (hep-th/9303057).

\define\RTSEYTLIN
A. Tseytlin, Phys. Lett. {\bf B242} (1990) 163;
Nucl. Phys. {\bf B350}
(1991) 395.

\define\RBOGOM
D. Olive and E. Witten, Phys. Lett. {\bf 78B} (1978) 97;
G. Gibbons and C. Hull, Phys. Lett. {\bf B109} (1982) 190.

\def\cm{{\cal M}}
\def\a{(\alpha)}
\def\cl{{\cal L}}
\def\p{\partial}
\def\aa{(a,\alpha)}
\def\bb{(b,\beta)}

{}~\vbox{\hbox{CALT-68-1879}
\hbox{hep-th/9307121}}\break

\title{DOES STRING THEORY HAVE A DUALITY
SYMMETRY\break RELATING WEAK AND STRONG COUPLING?
\foot{Supported in part by the U.S. Dept. of
Energy under Grant No. DE-FG03-92-ER40701.}}

\author{John H. Schwarz\foot{JHS@THEORY3.CALTECH.EDU}}

\address{California Insitute of Technology,
Pasadena, CA 91125}

\bigskip

\abstract

The heterotic string theory, compactified to four dimensions, has
been conjectured to have a duality symmetry
(S duality) that transforms the dilaton nonlinearly.
If valid,  this symmetry could
provide an important means of obtaining information about
nonperturbative features of the theory. Even though it is inherently
nonperturbative, S duality exhibits many
similarities with the well-established target-space duality symmetry
(T duality), which does act perturbatively. These
similarities are manifest in a new version of the low-energy
effective field theory and in the
soliton spectrum obtained by saturating the
Bogomol'nyi bound. Curiously, there is evidence that
the roles of the S and T dualities are interchanged in passing to a
five-brane formulation.

\vfill

\centerline{\it Presented at Strings `93 in Berkeley, California (May 1993)}

\endpage

\leftline{\bf Introduction}

This talk reports on
work done recently in collaboration with Ashoke Sen \RSSA \RSSB, which
investigated various issues concerning two types of duality
symmetries that have been considered in string theory. One,
which is well-established to all orders in string perturbation theory,
is known as ``target space duality,'' or  more succinctly as
``T duality.'' This discrete symmetry group is a generic feature of
theories with compactified spatial dimensions.
It is actually a discrete gauge group, so it should remain valid when
nonperturbative effects are taken into account. The simplest
example is the $Z_2$ symmetry that arises when one dimension
is compactified with the topology of a circle.
In this case, it can be described
as the equivalence of a circle of radius $R$ and one of radius
$\alpha'/R$. In fact, it can be realized as a
field transformation,
since $R$ corresponds to the classical value of a scalar field.
More generally, for the four-dimensional heterotic string,
compactified on a torus
that is dual to an even self-dual lattice in the manner proposed
by Narain \RNARAIN,
the corresponding T duality group turns out to be
O(6,22;Z). This example has N=4 supersymmetry, and is therefore
certainly unrealistic, but it is a particularly nice example to study.
Many, but not all,
of its properties are expected to hold in more realistic settings,
but that question will not be explored in detail here.

The second kind of duality, which is our main focus, is much more
speculative.  It is an SL(2,Z) group
that was discovered many years ago as a
symmetry of the classical field equations of
N=4 supergravity \ROLD. (This paper actually identified an SL(2,R)
symmetry. Instanton effects are
expected to break the symmetry to the discrete subgroup.)
N=4 supergravity contains a dilaton
$\phi$ and an axion $\chi$, which can be
combined in a complex scalar field $\lambda$ as follows
$$\lambda=\lambda_1+i\lambda_2 =\chi +i e^{-\phi} . \eqn\ethreetwo$$
This field transforms  nonlinearly under SL(2,Z)
$$\lambda\to {a\lambda + b\over c\lambda+d} ,\eqn\sltwo$$
where $a,b,c,d$ are integers satisfying $ad-bc=1$.
Since the value of the field $\lambda$ determines the
coupling constant $g$ and the vacuum angle $\theta$
according to
$$<\lambda> = {\theta \over 2\pi} + {8\pi i \over g^2} ,\eqn\vev$$
such a symmetry is necessarily nonperturbative. In particular,
the transformation $\lambda \to -1/\lambda$, for $\theta=0$,
inverts the coupling constant. The complex field $\lambda$ is
present in the massless spectrum even for compactifications that
only leave N=1 supersymmetry. It is therefore possible to speculate
that it might be a symmetry of the full nonperturbative
string theory
in that case, which is what was done in 1990 by Font {\it et al.}
\RFSSS.
This was, to the say the least, a very bold conjecture.
The proposed symmetry could be called ``dilaton--axion duality'' or
``weak coupling--strong coupling duality.''
These are rather cumbersome, so we
propose to refer to it as  ``S duality.'' While S duality is
still far from established, we find that all our studies support it.
Indeed, when analyzed in the proper way,
S duality and T duality have a great deal in common.

\medskip
\leftline{\bf 4D Effective Field Theory}

This section summarizes material I presented at Strings '92 in Rome \RSCHWARZ.
The low-energy effective field theory
that describes the massless bosonic fields associated with Narain
compactification of the heterotic string has a T
duality symmetry group $G_T$  = O(6,22;Z)  and scalar fields
(moduli) that parametrize the moduli
space O(6,22)/O(6)$\times$O(22)$\times G_T$. These fields are
conveniently described by a
28$\times$28 matrix-valued scalar field $M^{ab}$ satisfying the
constraints
$$M^T=M, \quad M^T L M=L, \eqn\ethreefive$$
where $L$ is the O(6,22) metric
$$L=\pmatrix{ 0 & I_6 & 0\cr I_6 & 0 & 0\cr 0 & 0 & -I_{16}}.
\eqn\ethreethree$$
For generic values of the moduli fields, the spectrum also contains
28 massless abelian gauge fields
(so that the gauge group is [U(1)]${}^{28}$).
For special values of the moduli there are
additional massless gauge fields and enhanced
gauge symmetry. However,  to keep things as simple as possible,
we will only include the 28 gauge
fields  $A_{\mu}^a$ that are massless for all values of the moduli.
The other massless bosons are the graviton
(described by a metric tensor $g_{\mu\nu}$), the dilaton $\phi$,
and an antisymmetric tensor
$B_{\mu\nu}$.  The action for this theory can be
obtained in a variety
of ways, one of the easiest of which is dimensional reduction from
ten dimensions \RMAHSCH. In terms of the ``string metric,'' the
result is
$$S = \int_M dx \sqrt{- g} ~ e^{- \phi}
({\cal L}_1 + {\cal L}_2 + {\cal L}_3 + {\cal L}_4+ {\cal L}_5)\ ,\eqn\sbn$$
$$\eqalign {{\cal L}_1 &= R \cr
 {\cal L}_2 &= g^{\mu \nu} \partial_\mu \phi \partial_\nu \phi\cr
 {\cal L}_3 &= - {1 \over 12} H_{\mu \nu \rho} H^{\mu \nu \rho}\cr
 {\cal L}_4 &= {1\over 8} g^{\mu\nu} {\rm tr} \big(\partial_{\mu}
M L \partial_{\nu} ML\big)\cr
 {\cal L}_5 &= -{1\over 4} F^a_{\mu\nu}(LML)_{ab}
F^{b\mu\nu} , \cr }\eqn\sbo$$
where
$${F}^a_{\mu \nu}  = \partial_\mu {A}^a_\nu - \partial_\nu
{A}^a_\mu \,\, .\eqn\sby$$
$$H_{\mu \nu \rho} = \partial_\mu B_{\nu \rho} + {1 \over 2} {A}^a_\mu
L_{ab} {F}^b_{\nu \rho} + ({\rm cyc.~ perms.})\,\,.\eqn\sbz$$
This result has manifest T duality since
the metric and dilaton are invariant under
T transformations, and the gauge fields transform
by the vector representation of O(6,22).

This theory also has S duality symmetry, though this is not at all
apparent in the form given above. To exhibit this symmetry, it is
convenient to replace the string metric by the canonical metric
by means of the Weyl rescaling
$g_{\mu \nu} \to e^{\phi}g_{\mu \nu}$, since the canonical metric
will be invariant under S duality, but the dilaton field is not.
Also, to exhibit the axion $\chi$, it is necessary to make a duality
transformation
$$ \sqrt{ -g} ~ e^{-2 \phi} H^{\mu \nu \rho} \to \epsilon^{\mu
\nu \rho \lambda} \partial_{\lambda}  \chi, \eqn\axion $$
which (as usual) interchanges the role of a field equation and a
Bianchi identity. Then one can introduce the complex field
$\lambda$ defined in eq. \ethreetwo\ and write a
``dual action,'' whose classical field equations are
equivalent to those
obtained from the original action.
$$S_{\rm dual} = \int_M dx \sqrt{- g}
({\cal L}_1^{\prime} + {\cal L}_{2,3}^{\prime}
+ {\cal L}_4^{\prime}+ {\cal L}_5^{\prime})\ ,\eqn\sbn$$
$$\eqalign {{\cal L}_1^{\prime} &= R \cr
{\cal L}_{2,3}^{\prime} &= -{1 \over 2\lambda_2^2}g^{\mu \nu}
\partial_\mu \lambda \partial_\nu \bar\lambda\cr
{\cal L}_4 ^{\prime}&= {1\over 8} g^{\mu\nu} {\rm tr} \big(\partial_{\mu}
M L \partial_{\nu} ML\big)\cr
{\cal L}_5^{\prime} &= - {\lambda_2\over 4}
F^a_{\mu\nu}(LML)_{ab} F^{b\mu\nu}+
{\lambda_1\over 4} F^a_{\mu\nu}L_{ab} \tilde F^{b\mu\nu} , \cr }\eqn\sbo$$
where
$$ \tilde F^{a\mu\nu}
={1\over 2 \sqrt{- g}}\epsilon^{\mu\nu\rho\sigma}
F^a_{\rho\sigma}.\eqn\ethreefour$$
All terms in the action $S_{\rm dual}$ are invariant under S
duality, except for ${\cal L}_5^{\prime}$.  However,
the equations of
motion do transform covariantly under S duality provided
that when $\lambda\to {a\lambda + b\over c\lambda+d}$
(and $ad - bc=1$),
$$ F^a_{\mu\nu}
\to c\lambda_2 (ML)_{ab}\tilde F^b_{\mu\nu} +(c\lambda_1+d)
F^a_{\mu\nu}. \eqn\ethreeseven$$
Note that in terms of the gauge fields themselves this is a nasty
nonlocal transformation.

\medskip
\leftline{\bf Manifest S Duality}

The construction of the action $S_{\rm dual}$ is a significant step towards
exhibiting S duality, but because of the noninvariance
of $\cl_5^{\prime}$, it is not the final form. We would like to
have a third form of the action with both dualities (S and T)
made manifest, thereby putting them on an equal footing.
The main problem in realizing S duality
in $S_{\rm dual}$ is attributable to the gauge fields. The basic idea
for overcoming this difficulty is to replace each gauge
field $A_{\mu}$
by a pair of independent gauge fields $A_{\mu}^{(\alpha)}$,
$\alpha = 1,2$ and to obtain the relation $F_{\mu\nu}^{(2)}
= \tilde F_{\mu\nu}^{(1)}$ as an equation of motion. The
formulas will not have manifest Lorentz invariance,
though they will
have manifest rotational symmetry. Accordingly, it is
convenient to
introduce separate ``electric'' and ``magnetic'' fields
$$E_i^{\a}=\p_0 A_i^{\a}-\p_i A_0^{\a}, \quad B^{\a i}=\epsilon^{ijk} \p_j
A_k^{\a} \quad 1\le i, j, k \le 3 . \eqn\etwotwo$$

Let us first describe a two-potential version  of
free Maxwell theory, and explain how that theory is coupled to gravity,
before applying the results to the problem of making S duality
manifest. Our action for free Maxwell theory is
$$ S=-{1\over 2}\int d^4x \Big(B^{\a i}\cl_{\alpha\beta}
E_i^{(\beta)}
+B^{\a i} B^{\a i}\Big),
\eqn\etwoone$$
where
$$\cl=\pmatrix{0 & 1\cr -1 & 0\cr}.
\eqn\etwothree$$
is the metric for SL(2,R) = Sp(2). The symbols ${\cal L}$
and ${\cal M}$ (introduced below) to describe S duality
are chosen to emphasize the analogy with $L$ and $M$ used in the description
of T duality. The action \etwoone\ has the following gauge
invariances
$$\delta A_0^{\a}=\Psi^{\a}, \quad \delta A_i^{\a}
=\p_i \Lambda^{\a}.\eqn\etwofour$$
It is easy to show that it describes a single propagating photon
with two physical polarizations. To understand how it works, note
that up to total derivatives the first term is proportional to
$\epsilon^{ijk}\p_0A_i^{(1)}\p_jA_k^{(2)}.$  Therefore the
fields $A_0^{(\alpha)}$ do not give any classical
equations of motion. Cast in this
form, the fields $A_i^{(2)}$ have no time derivatives and can be
treated as auxiliary. Thus the (integrated) equation of motion $B_i^{(2)}
=E_i^{(1)}$ can be used to eliminate $A^{(2)}_i$ from the action.
This gives rise to the standard Maxwell action in the
$A_0^{(1)}=0$
gauge. Gauss's law ($\p_i E_i^{(1)}=0$) is implied
by the Bianchi identity for $B_i^{(2)}$.
The action \etwoone\ is
manifestly invariant under the electric--magnetic duality
symmetry
$$A^{\a}_\mu\to \cl_{\alpha\beta} A^{(\beta)}_\mu,
\eqn\etwoten$$
which corresponds to $F_{\mu\nu} \to \tilde F_{\mu\nu}$.
This is to be contrasted with the usual Maxwell action,
${1\over 2}\int (E^2 - B^2) d^4 x$, which goes to its
negative. While not {\it manifestly} Lorentz invariant, the
action \etwoone\ is in fact invariant
under the global transformations
$$\delta A_i^{(\alpha)}= x^0 v^k \p_k A_i^{(\alpha)}
+\vec v \cdot \vec x \cl_{\alpha\beta}
\epsilon ^{ijk} \p_j A_k^{(\beta)}. \eqn \lorentz$$
On the mass shell,
this is identical to the usual Lorentz transformation formula
with boost parameter $v^i$.

Let us now consider how to couple \etwoone\ to gravity. Since we
do not have manifest Lorentz invariance before coupling to
gravity, we do not expect manifest general coordinate
invariance after
coupling to gravity. Nonetheless it is not difficult to figure out
which action gives the desired generally covariant field equations.
The result is
$$S_g=-{1\over 2}\int d^4 x \Big[ B^{\a i}\cl_{\alpha\beta} E^{(\beta)}_i
-{g_{ij}\over \sqrt{- g} g^{00}} B^{\a i} B^{\a j}+\epsilon^{ijk}
{g^{0k}\over g^{00}} B^{\a i}\cl_{\alpha\beta} B^{(\beta) j}\Big].
\eqn\etwofourteen$$
As usual, $\sqrt{-g} = \sqrt{-\det(g_{\mu\nu})}$
and $g^{\mu\nu}$ is the
inverse of $g_{\mu\nu}$, the ordinary four-dimensional metric.
These conventions are retained even when space and
time components are enumerated separately.
If one eliminates the fields $A_{\mu}^{(2)}$
by using their field equations as before, one
obtains the standard action $-{1\over 4} \int d^4 x \sqrt{- g}
F^{(1)\mu\nu} F^{(1)}_{\mu\nu}.$  Equation \etwofourteen\
is invariant under general coordinate transformations with
the metric transforming in the standard way and
$$\delta A_i^{\a}=\xi^j\p_j A_i^{\a}+(\p_i\xi^j) A^{\a}_j
+\xi^0\Big\{-{g_{ij}\over \sqrt{- g} g^{00}}\cl_{\alpha\beta}
B^{(\beta)j}-{g^{0k}\over g^{00}}\epsilon^{ijk} B^{\a j}\Big\}.
\eqn\etwofifteen$$
As in the case of Lorentz transformations, which corresponds to setting
$\xi^0 = \vec v \cdot \vec x$, $\xi^i = v^i x^0$, and $g_{\mu\nu}=
\eta_{\mu\nu}$, this differs from the
standard transformation formula by an amount that vanishes
when the classical equations of motion are satisfied.

Let us now return to the case of flat space-time and consider the
generalization of \etwoone\ that includes the coupling to the
axion--dilaton field $\lambda$. The appropriate formula is
$$S_\lambda=-{1\over 2}\int d^4 x[B^{\a i}\cl_{\alpha\beta} E^{(\beta)}_i
+B^{\a i} (\cl^T\cm \cl)_{\alpha\beta}B^{(\beta) i}],
\eqn\ethreeten$$
where
$$\cm(\lambda) ={1\over \lambda_2}
\pmatrix{1 & \lambda_1\cr \lambda_1 & |\lambda|^2\cr}. \eqn\ethreeeight$$
The matrix $\cm$ is a symmetric SL(2,R) matrix, which therefore satisfies
$\cm^T=\cm, \quad \cm\cl\cm^T=\cl$,
and $\cl$ is given in eq. \etwothree.
Under the SL(2,R) transformation \sltwo\ of the field $\lambda$,
$\cm\to \omega^T\cm\omega$, where
$\omega= \left({d\ b\atop c\ a}\right)$.
The action \ethreeten\ is manifestly invariant provided that at the same time
$A^{\a}_i\to (\omega^T)_{\alpha\beta} A^{(\beta)}_i$.
If we eliminate $A_i^{(2)}$ from \ethreeten\
by using its clasical equation of motion,
then we obtain the covariant expression
$\cl_5^{\prime}$ given in eq. \sbo.

We are now in a position to give a third version of our theory
that is
classically equivalent to the two versions given earlier, but with
both S duality and T duality realized as manifest symmetries. The
formula that does the job is
$$\eqalign{
S = \int  d^4 x \Big[& \sqrt{- g}\big\{ R - {1\over 4} g^{\mu\nu}
tr(\p_\mu \cm \cl \p_\nu\cm \cl) + {1\over 8} g^{\mu\nu}
Tr(\p_\mu M L \p_\nu M L)\big\}\cr
&-{1\over 2}\Big\{ B^{\aa i}\cl_{\alpha \beta} L_{ab} E_i^{\bb}
+ \epsilon^{ijk}{g^{0k}\over g^{00}}
B^{\aa i}\cl_{\alpha\beta} L_{ab} B^{\bb j}\cr &
-{g_{ij}\over\sqrt{- g} g^{00}} B^{\aa i}
(\cl^T\cm\cl)_{\alpha\beta}
(LML)_{ab} B^{\bb j} \Big\}\Big] .\cr}
\eqn\ethreetwentytwo $$
In the above equation $Tr$ denotes trace over the
indices $a,b$ and $tr$ denotes trace over the indices $\alpha, \beta$.
In this expression
we have recast the kinetic term of the
$\lambda$ field ($\cl_{2,3}^{\prime}$ in eq. \sbo)
in terms of the matrix $\cm$.

Written in the form \ethreetwentytwo,
it is clear that the S and T duality are realized in
quite analogous ways in the low energy effective action,
despite their profound difference from the point of view
of compactified string theory. As we have seen, the price
for making both of them manifest is the
loss of manifest general coordinate invariance.
The first form we presented (with manifest
T duality) was
derived by dimensional reduction of the ten-dimensional N=1
supergravity theory containing a two-form potential
$B_{\mu\nu}$.
However, there is also a dual version of that theory in which
the two-form potential is replaced by its dual, which is a
six-form potential $B_{\mu_1\mu_2 \cdots \mu_6}$
\RDUALSUGRA. Dimensional reduction of this dual theory
gives a version of the four-dimensional theory that has manifest
S duality but not T
duality. This fact gives an interesting insight into the possible
significance of S duality.

The two-form potential couples in a
natural way to the world volume of the string and is therefore
the natural choice from that point of view. However,
in similar fashion
the six-form potential couples to the world volume of a
`five-brane,' a five-dimensional extended object discovered as
a soliton solution of the heterotic string \RSTROM \RDUFF.
It has been conjectured in these references that there is a dual
formulation of the heterotic string theory in which the
five-brane is fundamental and the string is the soliton.
It is an interesting possibility that
in five-brane perturbation theory
S duality is true order-by-order,
whereas T duality becomes a non-perturbative
symmetry. In other words, the roles of  the S and T dualities
are interchanged
in passing from strings to five-branes, a phenomenon that we call ``duality of
dualities.'' As evidence in support of this conjecture,
it was shown that S duality describes the interchange of
five-brane Kaluza--Klein modes and winding modes, just as
T duality does for the compactified string theory.

Before continuing with our main theme, let us pause to describe
briefly generalizations of the two-potential formalism
we have used for
recasting the Maxwell action. What we have done is to introduce
a pair of  independent unconstrained potentials whose
field strengths are dual to one another
as a consequence of the equations of motion. This construction is
easily generalized to $m$-forms in $d$ dimensions. Let $A$
denote an $m$-form potential and $F$ its $(m+1)$-form field strength, so that
$$F_{\mu_1\ldots \mu_{m+1}}=\p_{[\mu_1}
A_{\mu_2\ldots \mu_{m+1}]}, \quad
0\le \mu_l\le d-1.\eqn\etenone$$
Similarly, let $B$ denote a $(d-m-2)$-form potential and $G$
its $(d-m-1)$-form field strength. Then form the action
$$\eqalign{S =\int d^dx &
\Big\{{\epsilon^{i_1\ldots i_m j_1\ldots
j_{d-m-1}}\over m! (d-m-1)!}
F_{0 i_1\ldots i_m} G_{j_1\ldots j_{d-m-1}}+
{1\over 2\cdot (m+1)!}F_{i_1\ldots i_{m+1}}
F_{i_1\ldots i_{m+1}}\cr
&+{1\over 2\cdot (d-m-1)!}
G_{i_1\ldots i_{d-m-1}}
G_{i_1\ldots i_{d-m-1}}\Big\},
\qquad 1\le i_l, j_l\le d.\cr}\eqn\etenfive$$
This action is invariant under the following gauge
transformations:
$$\eqalign{
\delta A_{0 i_1\ldots i_{m-1}}
=&\Psi^{(1)}_{i_1\ldots i_{m-1}}, \quad
\delta B_{0 i_1\ldots i_{d-m-3}}=
\Psi^{(2)}_{i_1\ldots i_{d-m-3}}\cr
\delta A_{i_1\ldots i_m}
=&\p_{[i_1} \Lambda^{(1)}_{i_2\ldots i_m]}, \quad
\delta B_{i_1\ldots i_{d-m-2}}=\p_{[i_1}
\Lambda^{(2)}_{i_2\ldots i_{d-m-2}]} .\cr}\eqn\etenseven$$
The field equations of \etenfive\ imply that $G$ is dual to $F$.
These equations provide a natural generalization of
eqs. \etwoone\ and \etwofour, which correspond to the case
$d=4$ and $m=1$. When $d=4n+2$
and $m=2n$ it is possible to have a self-dual field strength
(which amounts to equating $F$ and $G$).  An example
of this occurs in type 2B supergravity in ten dimensions, which
contains a self-dual five-form field strength. In this case our
formulas reduce to ones considered
previously by Henneaux and Teitelboim \RHENN.
In the particular case of a self-dual boson
in two dimensions, they agree with those of ref. \RJACKIW. This
case will be utilized in the next section.
These formulas also can be used to reformulate N=1 supergravity
in ten dimensions in a version containing {\it both} the two-form
$B_{\mu\nu}$ and the six-form $B_{\mu_1\mu_2 \cdots \mu_6}$.
This requires a slight generalization of the formulas given above
to accommodate the Chern-Simons terms that are present.

\medskip
\leftline{\bf T-Duality Symmetric World-Sheet Theory}

Let us now consider the dynamics of strings propagating in the
presence of  arbitrary background values of the fields
in the low-energy effective action. Specifically, they
are the string metric $g_{\mu\nu}(x)$ (which includes the dilaton
as a factor), the two-form potential $B_{\mu\nu}(x)$,
the moduli $M^{ab}(x)$, and the 28 abelian gauge fields
$A_{\mu}^a(x)$.  As is well-known,
this world-sheet theory is conformally invariant when
the backgrounds satisfy the appropriate equations of motion.
The dynamical `fields' of the world-sheet theory are the
space-time coordinates $x^{\mu}(\sigma,\tau)$, $\mu = 0,1,2,3$
and 28 internal coordinates $y^{a}(\sigma,\tau)$,
$a=1,2,\cdots,28$,
describing six compactified right-movers and 22 compactified left-movers.
The $y^a$'s are periodically identified, {\it i.e.},
$y^a \sim y^a + 2 \pi$.
The fact that they are chiral bosons of the
world-sheet theory is reflected in their world-sheet
equations of motion
$$D_0 y^a = - (ML)^a_b D_1 y^b, \eqn\missing$$
where
$$	D_\alpha y^a = \partial_\alpha y^a + A_\mu^a
	\partial_\alpha x^\mu. \eqn\twofour$$
Note that $(ML)^2 =1$, and the matrix $ML$ has 22 eigenvalues that are $-1$
and 6 eigenvalues that are $+1$. In addition, the
$x^{\mu}$ equations
of motion are
$$	\eqalign{g_{\mu\nu} \partial^\alpha \partial_\alpha x^\nu +
\Gamma_{\mu\nu\rho} \partial^\alpha x^\nu
\partial_\alpha x^\rho
	= &\ - {1\over 2} D_1 y^a (L \partial_\mu M L)_{ab} D_1 y^b
- \epsilon^{\alpha\beta} \partial_\alpha x^\nu F_{\mu\nu}^a
L_{ab} D_\beta y^b \cr & \quad
+ {1\over 2} \epsilon^{\alpha\beta} H_{\mu\nu\rho}
\partial_\alpha x^\nu \partial_\beta x^\rho ,\cr} \eqn\twothree$$
where $\Gamma$ is the usual  Christoffel symbol and $H$ is
defined in eq. \sbz. These equations are written in a form having
manifest  O(6,22,Z) symmetry (T duality). The restriction to
integers arises from the periodicity properties of the $y^a$.

At this point it is natural to wonder
whether these T duality symmetric equations can be obtained from a world-sheet
action that has this symmetry. This is certainly not true for the
usual formulation. In fact, it is easy to write down
such an action. The answer is
$$	\eqalign{S = {1\over 4\pi}
\int d^2\sigma \ & \Big\{g_{\mu\nu} \eta^{\alpha\beta}
\partial_\alpha x^\mu \partial_\beta x^\nu
-  D_0 y^a L_{ab} D_1 y^b -  D_1 y^a (LML)_{ab} D_1 y^b\cr
&+  \epsilon^{\alpha\beta} [B_{\mu\nu}
\partial_\alpha x^\mu \partial_\beta x^\nu - A_\mu^a
\partial_\alpha
x^\mu L_{ab} D_\beta y^b] \Big\} . \cr} \eqn\twofive$$
This result contains chiral bosons in the manner discussed earlier.
It is a generalization of a result given previously by Tseytlin \RTSEYTLIN.

To understand this theory better, it is important to exhibit the
coupling to a world-sheet metric $h_{\alpha\beta}$ that
gives 2D Weyl
invariance and reparametrization invariance.
This is achieved by
replacing the first term (as usual) by
$	 \sqrt{-h}  h^{\alpha\beta} g_{\mu\nu}(x)
\partial_\alpha x^\mu \partial_\beta x^\nu$,
and the third term by
$$	 {1\over \sqrt{-h} h^{00}} D_1 y^a (LML)_{ab}
D_1 y^b + {h^{01}\over  h^{00}} D_1 y^a L_{ab} D_1 y^b, \eqn\twoseven$$
which is analogous to the $B^2$ terms in eq. \ethreetwentytwo.
Having this form of the world-sheet action, it is straightforward to
deduce the (traceless) energy-momentum tensor
$T_{\alpha\beta}$
and the corresponding Virasoro constraints. It is also possible to
eliminate the world sheet metric to obtain a ``Nambu form'' that
maintains all the symmetries (including T duality).

The existence of the version of the world-sheet theory
presented above is actually quite remarkable.
It has manifest T duality, which is a symmetry that relates Kaluza--Klein
excitations of the string (which can be regarded as elementary
`particles' of the world-sheet theory) to winding-mode
excitations (which are solitons of the world-sheet theory). In terms of a
compactification scale $R$ and the string scale $\alpha'$, the
corresponding  masses are $M_{\rm KK} \sim 1/R$ and
$M_{\rm winding} \sim R/\alpha'$.
Thus T duality is a nonperturbative symmetry of the world-sheet theory,
just as S duality is conjectured to be for the space-time theory.
A world-sheet theory that makes
such a nonperturbative symmetry manifest
is necessarily strongly coupled. We know it is correct, however,
since the $y$ coordinates only appear quadratically and can
therefore be treated exactly. The analogy with the space-time theory
raises the very interesting question whether it is possible to formulate
a string field theory with manifest S duality. (It is already
known how to implement T duality in string field theory \RKUGO.)

\medskip
\leftline{\bf Discussion}

The toroidally compactified heterotic string has an infinite
spectrum of
excitations carrying electric and magnetic charges (with respect to
the 28 abelian gauge fields).  The states that carry electric charges
only are elementary in the sense that they have a perturbative
description in the space-time theory. States that carry magnetic
charge, on the other hand, must be regarded as solitons of the
string theory. The electric and magnetic
charges of a state can be defined
by the asymptotic behavior of the gauge fields
$$F^a_{0i} \sim q_{\rm el}^a\, {x^i\over r^3} \quad \quad
\tilde F^a_{0i} \sim q_{\rm mag}^a \,{x^i\over r^3} .
\eqn \elmag$$
The allowed values of the electric and magnetic charges are
determined by the asymptotic values
of the moduli fields ($M_{ab}^{(0)}$ and $\lambda^{(0)}$).  In terms of these
one has \RDYONO
$$q_{\rm el}^a={1\over \lambda_2^{(0)}}M^{(0)}_{ab}
(\alpha_0^b+\lambda_1^{(0)}\beta_0^b), \quad
q_{\rm mag}^a=L_{ab}\beta_0^b,
\eqn\ebbthree$$
where both $\alpha_0^a$ and $\beta_0^a$ are 28-component
vectors
belonging to a reference lattice $P_0$, which is
even and self-dual with respect to the metric $L$. These charges
automatically incorporate the Dirac quantization condition
suitably generalized to allow for dyons and a vacuum
angle \RDIRAC. The appearance of the electric and
magnetic charges as central charges in the
supersymmetry algebra allows one to deduce a
Bogomol'nyi lower bound on the masses of states with specified
electric and magnetic charges \RBOGOM. In the present
context this bound is given by
$$(m_0)^2={1\over 16} \pmatrix{\alpha_0^a & \beta_0^a\cr}
\cm^{(0)} (M^{(0)}+L)_{ab} \pmatrix{\alpha_0^b\cr \beta_0^b \cr}.
\eqn\ebogom$$
Remarkably, this formula turns out to be symmetric under both
S and T dualities, providing yet further evidence that at a
fundamental level they should operate in much the same way.

The perturbative string spectrum contains all states with electric
charges only, in other words all states with $\beta_0^a =0$. This
is a T duality invariant
(but not S duality invariant) subset of
the complete spectrum. The perturbative five-brane spectrum, on the
other hand, contains all states for which the last 22 components
of $\alpha_0^a$ and $\beta_0^a$ vanish. This is an S duality
invariant (but not T duality invariant)
subset of the complete spectrum. In view of these
facts, it is tempting to speculate that the
classical five-brane theory has S duality symmetry much
as the string world-sheet theory has T duality symmetry.
However, the severe
nonlinearities of the five-brane theory have so far prevented us
from proving this.

Since the S duality group SL(2,Z) relates electrically
charged states to magnetically charged states, it relates
perturbative
states and nonperturbative states of the space-time theory,
just as the
T duality group O(6,22;Z) did for the world-sheet theory.
Specifically, the SL(2,Z) group element
$\left({a\ b\atop c\ d}\right)$ maps states
with charges $(\vec\alpha_0,0)$ to ones with charges
$(a\vec\alpha_0, c \vec\alpha_0)$.
It is possible to find group elements for any pair of relatively prime
integers $a$ and $c$. If we simultaneously allow the
transformations
to act on the background fields of the world-sheet action to
give transformed background fields, whose relationship to the
original ones are nonlocal in general, then we obtain a ``dual'' world-sheet
action that is isomorphic to the original one. From the point of
view of this dual theory states with charges of the form
$(a\vec\alpha_0, c \vec\alpha_0)$ arise perturbatively
and all others
are solitons. Thus we have a generalization of Olive--Montonen
duality \ROLIVE\ to a situation where there are an
infinite number
of isomorphic dual descriptions of the same theory labeled by
pairs of relatively prime integers.

To conclude, we have seen that S duality and T duality have many
similarities. To the extent that a fundamental
five-brane formulation
of the heterotic string theory
makes sense, S duality should be a fundamental symmetry.
However, even if five-branes turn out to only make sense as
solitons, the symmetry could still be true.

I wish to acknowledge the hospitality of the Institute of
Theoretical
Physics  at U.C. Santa Barbara, where this work was done.

\refout

\end